\documentclass[%
 preprint, 
 amsmath,amssymb,
 aps, physrev,
]{revtex4-2}

\usepackage{graphicx}
\usepackage{dcolumn}
\usepackage{bm}
\usepackage[T1]{fontenc}
\usepackage[utf8]{inputenc}
\usepackage{booktabs}
\usepackage{physics}
\usepackage{siunitx}
\usepackage{multirow}
\usepackage{ulem}
\usepackage{bm}
\usepackage{float}
\usepackage{placeins}

\usepackage[draft]{changes}
\definechangesauthor[name=PDD, color=red]{3}

\begin{document}

\title{Source function from two-particle correlation function through entropy-regularized Richardson-Lucy deblurring}

\author{Chi-Kin Tam}
 \email{Contact author: chikin.tam@wmich.edu}
 \affiliation{Department of Physics, Western Michigan University, Kalamazoo, Michigan 49008, USA
 }%
\author{Zbigniew Chaj\k{e}cki}
 \affiliation{Department of Physics, Western Michigan University, Kalamazoo, Michigan 49008, USA
 }%
\author{Pawe\l{} Danielewicz}
 \affiliation{Facility for Rare Isotope Beams and Department of Physics and Astronomy, Michigan State University, East Lansing, Michigan, USA
 }%
\author{Pierre Nzabahimana}
 \affiliation{Facility for Rare Isotope Beams and Department of Physics and Astronomy, Michigan State University, East Lansing, Michigan, USA, 
 }%
 \affiliation{Los Alamos National Laboratory, Los Alamos, NM 87545, USA
 }%

\date{\today}

\begin{abstract}
Source functions are obtained from $p$-$p$ and $d$--$\alpha$ correlation functions by applying the Richardson-Lucy (RL) deblurring to the Koonin-Pratt (KP) equation. To prevent fitting of noise in the correlation function, total-variation (TV) regularization is employed that has been  effective in ordinary image restoration. TV alone cannot ensure normalization of the source functions. To ensure the latter, we propose a maximum-entropy regularized RL algorithm (MEM-RL). We outline the MEM-RL formalism and optimization strategy for the KP equation, demonstrating its effectiveness on both simulated and experimental data, including the $p$-$p$ and $d$--$\alpha$ correlation functions.
\end{abstract}
\maketitle

\section{Introduction}

Over the years, the two-particle correlation functions have been extensively studied to access the spacetime extent of the emission source and probe particle emission-mechanisms~\cite{fang2016emission} in heavy--ion collisions over a wide range of energies~\cite{giuseppe2001imaging, giuseppe2007dalpha, henzl2011hira, Wang:2021mrv}. In particular, the amplitude and shape of the two-proton correlation $C_{pp}(q)$, as measured across distinct gates on pair momentum, serve as indicators of emission-source sizes across various timescales~\cite{zhu1991}. Moreover, the particle correlations for $d$--$\alpha$, $p$--$\alpha$, and $\alpha$--$\alpha$ exhibit characteristic peaks originating from the resonance decays of their associated excited states, providing a unique opportunity to investigate the structure of compound systems and interactions with the particle pair.

The amplitude and the shape of the correlation function depend on the interaction between the particle species in the pair. For identical pairs such as two protons, the broad peak centered at $q \approx 20~\mathrm{MeV}/c$ arises from the S-wave final-state interaction. The interplay between strong-interaction attraction, Coulomb repulsion, and Bose symmetrization determines the shape of the correlation function at lower momenta. For other particle pairs, the correlation function also ties to the resonance decays. For instance, $d$--$\alpha$ pairs are primarily produced from the resonance decay of the excited states of $^6\mathrm{Li}^*$ at $2.186$ MeV and $4.312$ MeV, giving rise to two strong peaks~\cite{giuseppe2007dalpha}. Under the smoothness assumption, i.e., the single particle source varying slowly with the momentum~\cite{pratt1997smoothness}, the correlation functions get connected to the underlying source function $S(r)$ through the angle-averaged Koonin-Pratt equation~\cite{konnin1977, pratt1990kp, pratt1987kp, gong1991kp}. 
\begin{gather}
    C(q) = 1 + 4 \pi \int dr\, r^2 S(r) K(q, r),
    \label{eq:KooninPratt}
\end{gather}
where $q = \mu|\vec{\mathbf{v}}_1 - \vec{\mathbf{v}}_2| / 2$  and $r = |\vec{\mathbf{r}}_1 - \vec{\mathbf{r}}_2|$ refer to the relative momentum and the relative distance of the pair with $\vec{\mathbf{v}}_{1,2}$ and $\vec{\mathbf{r}}_{1,2}$ the velocities and positions of the two particles, $1$ and $2$. The kernel $K(q, r)$, defined as the squared wavefunction of the pair subtracted from unity, $|\psi(q,r)|^2 - 1$, encodes all physical consequences of the interactions. It is usually derived by solving the Schr\"odinger equation for scattering potential determined from matching the asymptotic behavior of the wavefunction to the measured phase shifts~\cite{messiah1961quantum}. The source function $S(r)$ represents the probability of emission of a pair at a relative distance $r$ at the time when the second particle is emitted. It is often normalized to a purity parameter $\lambda \le 1$ to account for the fact that a fraction of particles are emitted from processes over long timescales, such as evaporation and secondary decays. Thus, not every pair of particles is correlated with each other. 

Experimentally, the correlation function in Eq.~\eqref{eq:expcf} is defined as the ratio of spectra of measured pairs in the same event $A(\vec{\mathbf{p}}_1,\vec{\mathbf{p}}_2)$ to an uncorrelated background,
$B(\vec{\mathbf{p}}_1,\vec{\mathbf{p}}_2)$, where $\vec{\mathbf{p}}_{1,2}$ are the momenta of the two particles, 1 and 2:
\begin{gather}
    C(q) \propto  \dfrac{A(\vec{\mathbf{p}}_1,\vec{\mathbf{p}}_2)}{B(\vec{\mathbf{p}}_1,\vec{\mathbf{p}}_2)}.
    \label{eq:expcf}
\end{gather}
The background is usually constructed by the event-mixing method~\cite{kopylov1974,lisa1991eventmixing}, in which the particles are selected from different events with identical single-particle phase-space. The normalization is determined by either the number of pairs or explicitly set to unity at large relative momentum $q$, i.e., $C(q\rightarrow \infty) = 1$, assuming no correlations of any type at large~$q$. Our goal is to infer the unknown source function $S(r)$ from the measured correlation function $C(q)$ given the full knowledge of the kernel $K(q,r)$ through the KP equation. This constitutes a non-blind deconvolution problem. 

In the literature, the methods for restoring the source function can be categorized into three types. The first one is to estimate the source with a parameterized form of the source function where the parameters are optimized according to a $\chi^2$ fit. It has been found that $C_{pp}(q)$ from the central collision at $50$ MeV/nucleon can be well described, assuming the source function to be a Gaussian distribution~\cite{henzl2011hira}. Despite its effectiveness and simplicity, the method is inferior since any fine structure of the source function deviating from the chosen parameterization cannot be inferred. The second method, known as imaging~\cite{dave2001imaging, dave2005imaging3d}, solves the inverse problem by sampling the probability density of the source function given the correlation function. For Gaussian likelihood, the most probable distribution coincides with the one that minimizes the $\chi^2$ of the data. Imaging has been extensively applied to reconstruct source functions from various systems~\cite{giuseppe2001imaging, henzl2011hira, panitkin2001imaging, chung2003imaging} and to extract higher-order moments~\cite{dave2005imaging3d}. The solution from imaging is stabilized by explicitly introducing constraints such as a zero slope at the origin and the vanishing source at the endpoint.

Finally, the Richardson-Lucy (RL) algorithm~\cite{richardson1972, lucy1974}, originally developed in the fields of optics and astronomy, searches the source functions in an alternative Bayesian approach. Consider an inverse problem without noise, the measured distribution of an quantity $t$ is given by the convolution of the true distribution of the underlying object $t'$ and a blurring kernel $A$. Mathematically, 
\begin{gather}
    \label{eq:integral-rl}
    y(t) = \int A(t|t') x(t') \, \dd t' \, ,
\end{gather}
where $A(t|\tilde{t})\,\dd t$ represents the probability of measurement in the interval $(t, t+dt)$ given the truth $t'=\tilde{t}$. For discrete measurements due to the finite energy resolution of detectors, Eq.~\ref{eq:integral-rl} is rewritten in matrix form
\begin{gather}
    \label{eq:forward-model}
    \mathbf{y} = \mathcal{A}\mathbf{x},
\end{gather}
Here, $\mathbf{y}$ and $\mathbf{x}$ refer to the data and underlying object and the forward operator $\mathcal{A}$ are normalized by the definition of probability. In such case, the RL algorithm iteratively updates the object $\mathbf{x}$, which in the present case is the source function, by
\begin{gather}
    \label{eq:standard-rl}
    \mathbf{x}^{(r+1)} = \mathbf{x}^{(r)} \bigg[\mathcal{A}^T  \dfrac{\mathbf{y}}{\mathbf{y}^{(r)}}\bigg],
\end{gather}
where $r$ refers to the number of iterations and $\mathbf{y}^{(r)} = \mathcal{A}\mathbf{x}^{(r)}$ is the $r^\text{th}$ estimation of the observed data. 

Starting from an initial guess, typically a uniform function, the algorithm updates the source function multiplicatively and increases the likelihood of the observed samples at each iteration. One can see from equation Eq.~\ref{eq:standard-rl} that the integral of the object is conserved, $\sum x_i^{(r+1)} = \sum x_i^{(r)}$, and stays non-negative $x_i \ge 0 \, \forall i$ at all iterations, provided that the initial guess $\mathbf{x}^{(0)}$ satisfies the same conditions. These features are perfect for counting experiments due to the positive nature of the data. In astronomy and astrophysics, the algorithm has been applied to deblurring images taken with X-ray and $\gamma$-ray telescopes~\cite{sakai2023deblur, atwood2009deblur, prato2012deblur} and deprojection problem~\cite{konrad2013deblur}. In nuclear physics, the algorithm has been employed to perform invariant-mass spectra deconvolution~\cite{pierre2022deblurring} and to reconstruct reaction plane angle in heavy ion collision~\cite{pawel2022deblurring}. A coarse structure of the object will be essentially recovered in a few iterations, and a small deviation of prediction $\mathbf{y}^{(r)}$ from data $\mathbf{y}$ will be averaged out when convoluted with $\mathcal{A}$. 

The Koonin-Pratt equation shares the same mathematical structure as Eq.~\ref{eq:forward-model} and lends itself to the application of RL algorithm Eq.~\ref{eq:standard-rl}. To stabilize the solution, one has to introduce regularization to the RL algorithm to prevent over-fitting. Recently, a source function associated with $d$--$\alpha$ correlations in $^{40}\mathrm{Al} + ^{27}\mathrm{Al}$ at $44$ $\mathrm{MeV}/\mathrm{A}$ has been successfully restored using the RL algorithm with total-variation (TV) regularization~\cite{pierre2023deblurring}, which is effective in preserving sharp edges and is popular in fields of the image reconstruction. 

The TV regularization is derived by adding a penalty term to the objective function that is proportional to the sum of the absolute value of the gradient of the image~\cite{rudin1992tvreg}. To be explicit, it introduces an extra multiplicative factor $1/(1\mp \lambda_{\mathrm{TV}})$ in the update equation, smoothing the image in the entire domain by lowering the peaks and raising the valleys. However, it poses challenges in optimizing the regularization parameter $\lambda_{\mathrm{TV}}$. If $\lambda_{\mathrm{TV}}$ is too small, the smoothing effect is not strong enough to suppress unphysical features. While too large value of $\lambda_{\mathrm{TV}}$ would inevitably destroy the normalization of the source function, in which case the algorithm simply diverges.

In this work, we attempt to regularize the RL algorithm with the maximum-entropy method (MEM) proposed in the work~\cite{lucy1994} by Lucy, which naturally preserves the integral of the source function. The paper is organized as follows. In section~\ref{sec:formalism}, we first formulate the update rule using the standard RL algorithm with Koonin-Pratt equation as the forward model. Then, we introduce the modification of the update rule due to the maximum-entropy regularization and the associated choice of the default solution. In section~\ref{numerical-tests}, we examine the effectiveness of the algorithm on a correlation function associated with a known Gaussian source function and illustrate the optimization steps for the regularization parameters. In section~\ref{sec:result}, we apply the algorithm to experimental data for two-proton correlation function in~\cite{henzl2011hira} and to preliminary $d$--$\alpha$ correlation from a recent experiment~\cite{sweany2021hira}. Finally, we discuss the results and practical challenges.

\section{Formalism}
\label{sec:formalism}
\subsection{Koonin-Pratt equation}
Our objective is to restore the source function $S(r)$ in~\eqref{eq:KooninPratt} from measured correlation function $C(q)$ using a deblurring algorithm applicable to inverse problem of Eq.~\eqref{eq:forward-model}. To utilize the formalism in Eq.~\eqref{eq:standard-rl}, the Koonin-Pratt equation is expressed in a matrix form to recognize the forward model:
\begin{gather}
    \label{eq:kp-forward}
    \mathcal{C} = \mathcal{K} \mathcal{S}, \quad \text{with,} \\
            \label{K-matrix}
    \mathcal{K}_{ij} = 4\pi r_j^2 \bigg(K(q_i, r_j) + \dfrac{1}{\lambda} \bigg) \Delta r_j \, .
\end{gather}
Here, the purity parameter $\lambda$ is absorbed to the matrix $\mathcal{K}$ by using the normalization constraint $\lambda = 4\pi\int \dd r\, r^2S(r)$. The separation $r_j$ refers to the $j$'th bin of the relative distance, $q_i$ refers to the $i$'th bin of the relative momentum of a pair. Upon defining $\Delta q$ and $\Delta r$ as the bin widths of relative momentum and distance respectively, the discretized kernel can be expressed as
\begin{gather}
    K(q_i,r_j) = \dfrac{1}{\Delta q}\int_{q_i-\Delta q/2}^{q_i+\Delta q/2} \dd q \dfrac{1}{\Delta r} \int_{r_j-\Delta r/2}^{r_j+\Delta r/2} \dd r \, K(q,r) \, .
\end{gather}

\begin{figure}
    \centering
    \includegraphics[width=0.5\textwidth]{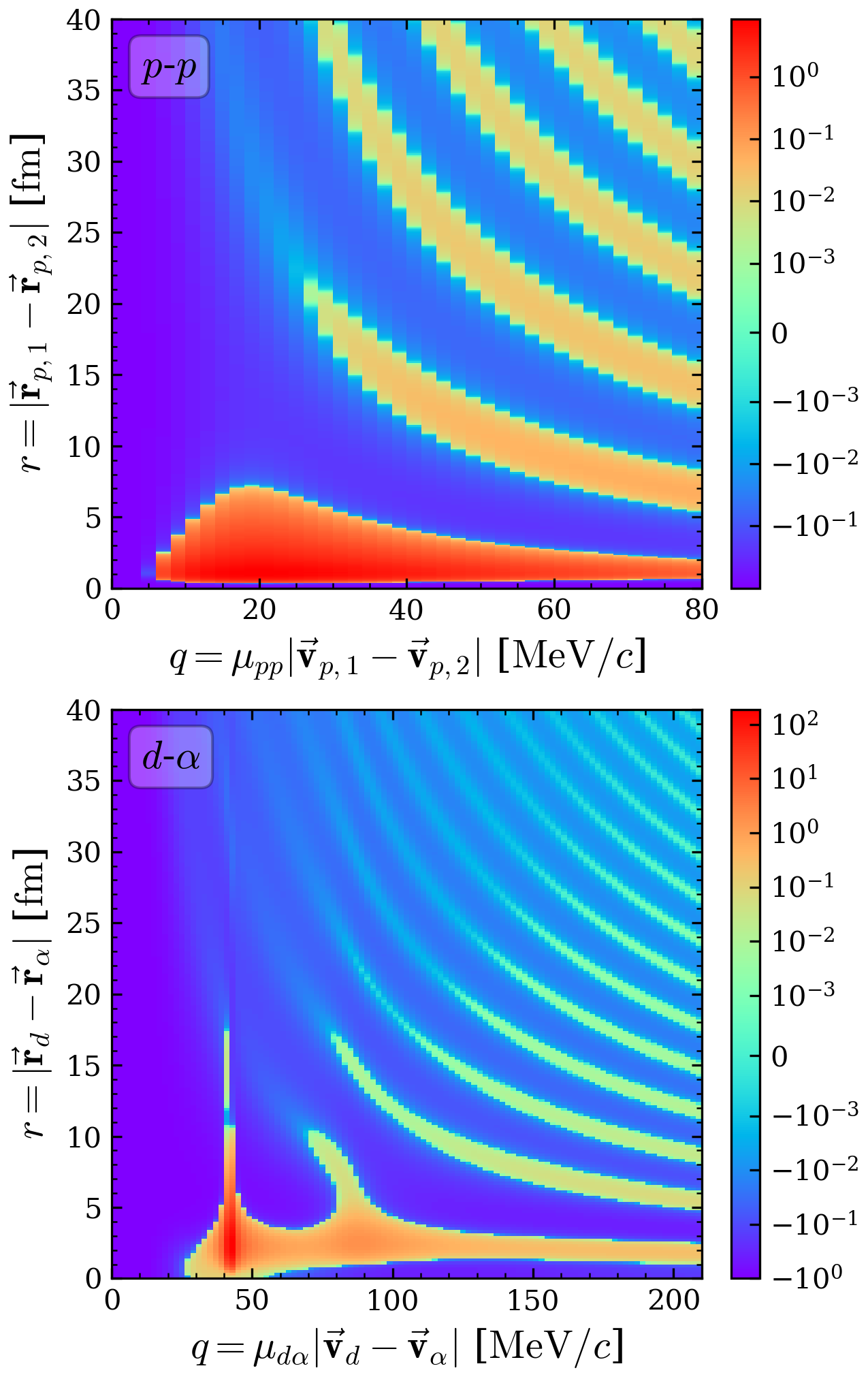}
    \caption{Kernels employed in the Koonin-Pratt equation plotted in a logarithmic scale. The upper panel shows the kernel for a two-proton pair with $N_q=40$ bins, $q \in (0,80) \mathrm{MeV}/c$. Lower panel shows the kernel for a deuteron-$\alpha$ pair with $N_q=50$, $q \in (0,200) \mathrm{MeV}/c$. Resonance peaks and Coulomb tails are clearly visible in both kernels. }
    \label{fig:kernel}
\end{figure}

\subsection{Richardson-Lucy algorithm}
Without loss of generality, all quantities in Eq.~\eqref{eq:kp-forward} are normalized with respect to the sum over their respective domains, 
\begin{align}
    \label{eq:normalization}
    \begin{split}
    \mathcal{C}_i &\rightarrow \mathcal{C}_i / \mathcal{C}^{(\mathrm{norm})} = \mathcal{C}_i / \sum_i \mathcal{C}_i, \\
    \mathcal{S}_j &\rightarrow \mathcal{S}_j / \mathcal{S}_j = \mathcal{S}^{(\mathrm{norm})} / \sum_j \mathcal{S}_j, \\
    \mathcal{K}_{ij} &\rightarrow \mathcal{K}_{ij} / \mathcal{K}^{(\mathrm{norm})}_j = \mathcal{K}_{ij} / \sum_i \mathcal{K}_{ij} \, ,
    \end{split}
\end{align}
to satisfy the requirement of Eq.~\eqref{eq:standard-rl}. Provided that the initial guess satisfy the normalization constraint $\sum_j\mathcal{S}^{(r=0)}_j = 1$, the physical source function $S(r)$ can be restored by scaling the converged output $\mathcal{S}$ by an normalization factors of $\mathcal{C}^{(\mathrm{norm})} / \mathcal{K}^{(\mathrm{norm})}$. From now on, the quantities $\mathcal{C}, \mathcal{S}$ and $\mathcal{K}$ refer to their normalized versions.

Upon comparing the functional form in Eq.~\eqref{eq:forward-model} and Eq.~\eqref{eq:kp-forward}, it is apparent that the standard RL algorithm Eq.~\eqref{eq:standard-rl} for deblurring the source function can be derived by identifying $\mathcal{C} \leftrightarrow \mathbf{y}$ and $\mathcal{S} \leftrightarrow \mathbf{x}$ and $\mathcal{K} \leftrightarrow \mathcal{A}$. In order to facilitate the derivation in later section, the iterative rule is rewritten in an additive form
\begin{gather}
    \label{eq:rl-update}
    \Delta^H \mathcal{S}_j^{(r)} = \mathcal{S}_j^{(r+1)} - \mathcal{S}_j^{(r)} = \mathcal{S}_j^{(r)} 
    \bigg[
        \sum_i \dfrac{\tilde{\mathcal{C}}_i}{\mathcal{C}_i^{(r)}}\mathcal{K}_{ij} - 1
    \bigg], 
\end{gather}
where the tilde sign in $\tilde{C}_i$ emphasizes the quantity is obtained from the data.

\subsection{Maximum-entropy regularization}
To prevent over-fitting the noise in the data, Lucy~\cite{lucy1994} demonstrated efficiency of a regularization scheme based on maximizing the entropy
\begin{gather}
    \mathbb{S} = -\sum_j \mathcal{S}_j\ln \dfrac{\mathcal{S}_j}{\chi_j},
\end{gather}
where $\chi$ refers to a prior solution. The bias of the estimate will be roughly minimized if the prior is taken to be an approximation of the truth. Since such a model does not exist, one can dynamically construct the prior as a smoothened solution at each iteration. This method, known as the floating default~\cite{lucy1994, 1985floating}, is built by defining 
\begin{gather}
    \label{eq:floating-default}
    \chi_j = \sum_k \Pi_{jk} \mathcal{S}_k,
\end{gather}
where $\Pi$ is any non-negative, properly normalized matrix of $[r_j, r_k]$ and is usually symmetric. We adopt the standard choice of $\Pi$ as a Gaussian-smearing kernel
\begin{gather}
    \label{eq:smearing-kernel}
    \Pi_{jk} \propto \exp\bigg(-\dfrac{(r_j-r_k)^2}{2\sigma_r^2}\bigg) \, .
\end{gather}
Lucy~\cite{lucy1994} showed that the regularization term can be added to the update rule of the RL algorithm as
\begin{gather}
    \label{eq:me-update}
    \Delta^\mathbb{S}\mathcal{S}_j = -\alpha \mathcal{S}_j \bigg[
        \mathbb{S} + \ln \dfrac{\mathcal{S}_j}{\chi_j} + 1 - \sum_k \dfrac{\mathcal{S}_k}{\chi_k}\Pi_{kj} \, ,
    \bigg],
\end{gather}
where the regularization strength $\alpha$ and smoothing width $\sigma_r$ are to be optimized. The method is advantageous over the TV regularization since the normalization constraint is strictly respected, i.e., $\sum_j \Delta^\mathbb{S}\mathcal{S}_j = 0$ and $\sum_j \Delta^\mathbb{H}\mathcal{S}_j = 0$ independently, and thus divergence is avoided. The method also achieves greater flexibility in the degree of smoothness through the functional form of $\Pi$, which comes at the cost of introducing additional optimization parameters.

\subsection{Accelerated convergence}
To achieve the convergence in a reasonable number of iterations, we introduce an acceleration coefficient $\nu$. The complete update rule of the MEM-RL algorithm is given by
\begin{gather}
    \label{eq:final-update}
    \Delta \mathcal{S}_j = \nu \bigg(\Delta^H \mathcal{S}_j + \Delta^\mathbb{S}\mathcal{S}_j\bigg) \, ,
\end{gather}
where $\Delta^H$ refers to the data-driven term in Eq.~\eqref{eq:rl-update} and $\Delta^\mathbb{S}$ refers to the regularization term in Eq.~\eqref{eq:me-update} which honors smoothness. The value of $\nu$ should be chosen such that a non-negativity constraint is not violated but is not too large that the algorithm fails to converge. In this work, $\nu = 1.99$ is generally a good choice. The complete procedure is as follows. For a given set of parameter ($\lambda$, $\alpha$ and $\sigma_r$), the $\lambda$-dependent $\mathcal{K}$ matrix is defined in~\eqref{K-matrix} and the smoothing matrix $\Pi(\sigma_r)$ is constructed according to Eq.~\eqref{eq:smearing-kernel}. The algorithm initializes with a normalized, uniform function $\mathcal{S}^{(0)}$. The $\mathcal{K}$ matrix and the data $\mathcal{C}$ are normalized according to Eq.~\eqref{eq:normalization}. With the prior constructed according to Eq.~\ref{eq:floating-default}, the output of the RL algorithm is then updated through Eq.~\eqref{eq:final-update} until the convergence is achieved. Then, the source function is restored by scaling the output using the appropriate normalization factors. The set of (hyper-)parameters is optimized by minimizing the $\chi^2$ of the associated correlation function to the data.

\section{Numerical Tests}
\label{numerical-tests}
We perform numerical tests by applying the standard and maximizing-entropy algorithm to correlation functions originating from the Gaussian source function parameterized as 
\begin{gather}
    \label{eq:gaus-source}
    S_G(r;\lambda_G, R_G) = \dfrac{\lambda_G}{(2\sqrt{\pi}R_G)^3} \exp\bigg(-\dfrac{r^2}{4R_G^2}\bigg).
\end{gather}

The correlation functions are generated by convoluting the sources with the kernel according to Eq.~\eqref{eq:kp-forward}. For simplicity, we primarily focus on the case of two-proton in which the correlation exhibits a single peak dominated by the strongly attractive singlet S-wave interaction~\cite{pawel2007coral,pratt2003coral}, as shown in the upper panel of Fig.~\ref{fig:kernel}.

\subsection{Test using the standard RL algorithm}
\label{sec:Test}
As a consistency check and to establish a baseline for MEM-RL, we apply the standard RL algorithm Eq.~\eqref{eq:standard-rl} to restore a source function from a noise-free correlation function. To ensure convergence to a unique solution, the dimension in data must not be smaller than that in the underlying source, i.e., $N\ge M$~\cite{lucy1974}. For a typical measurement of $C_{pp}(q)$, the data (and thus the $q$-axis of the kernel shown in Fig.~\ref{fig:kernel}) is usually binned with the width in~$q$ of $2-4 \mathrm{MeV}/c$, limited by the detector resolution and statistics. The binning of the kernel in the $r$-direction is non-trivial. It has to satisfy the dimensionality constraint but retain as much information as possible. That is, the associated correlation function should be a close approximation of the one obtained from the continuous source function. Due to $r^2$ dependence in the KP equation, the first bin must be chosen to be significantly away from the origin to avoid divergence. We have chosen $N=25$ and $M=13$ for~$q \in (0, 100)~\mathrm{MeV}/c$ and $r \in (0, 39)$~fm, respectively. Here, the kernel used in the process of restoration is identical to the one used in the generation of the correlation function. 
\begin{figure}
    \centering
    \includegraphics[width=0.6\textwidth]{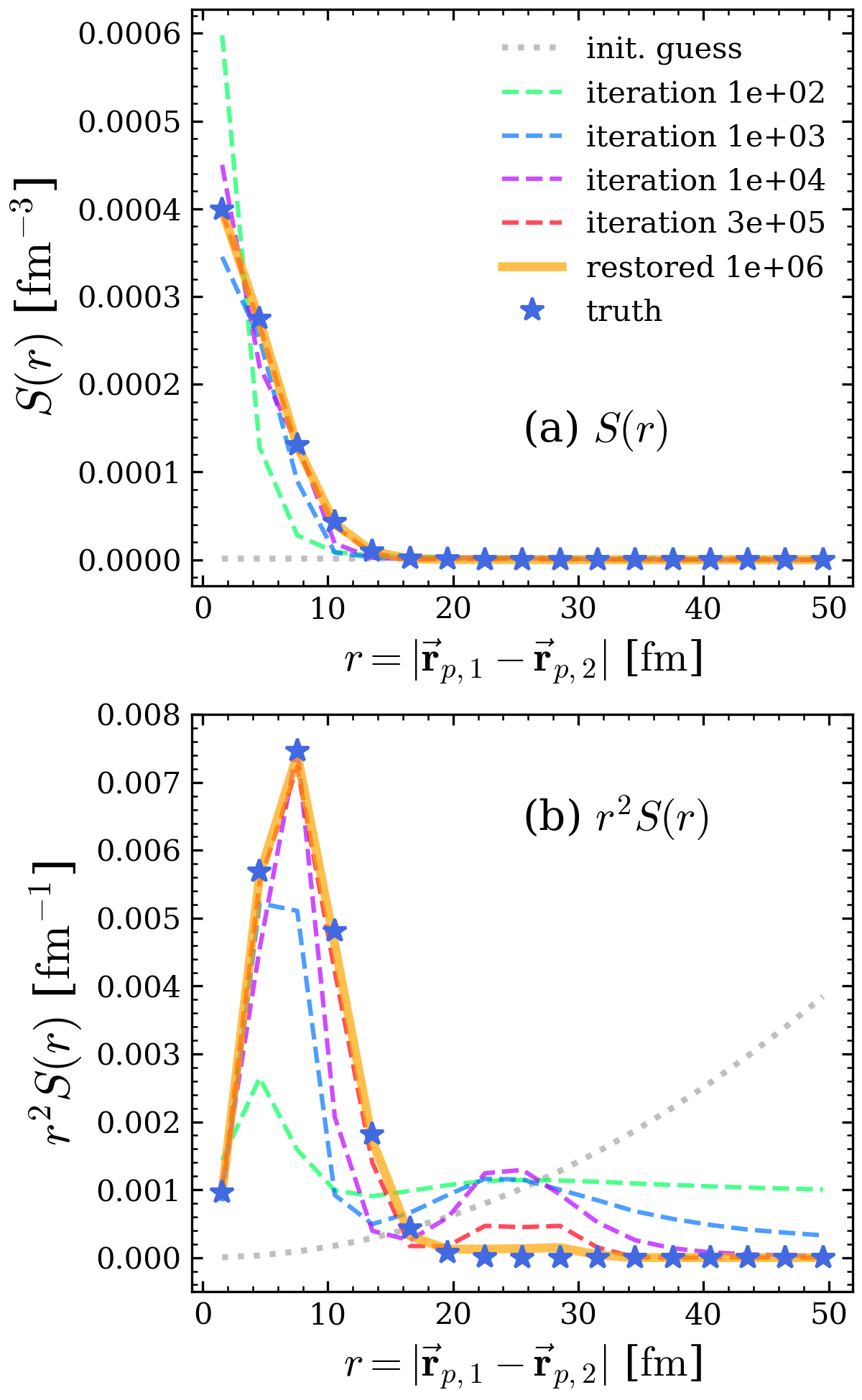}
    \caption{Restoration of the Gaussian source function from its associated smooth correlation function using the standard RL algorithm. The top panel shows the restored source function at different numbers of iterations. The bottom panel shows the restored source function scaled by $r^2$. The blue star represents the true source function; the gray dotted line represents the initial guess of the source function; dotted lines of different colors represent the restored source function at different numbers of iterations. The solid orange line represents the restored source function at convergence.}
    \label{fig:standard_rl_test}
\end{figure}
The restored source functions are shown in the top panel of Fig.~\ref{fig:standard_rl_test}. Uniform source functions (gray dotted line) with the same purity are used as the initial guess. The bottom panel of Fig.~\ref{fig:standard_rl_test} shows the $r^2S(r)$ distribution and is used to monitor any unphysical features of the source function at large $r$. While the overall structures are restored in a relatively small number of iterations, the algorithm requires significantly more iterations to achieve convergence, which is a known characteristic of the RL algorithm.

\begin{figure}[!t]
    \centering
    \includegraphics[width=1.0\textwidth]{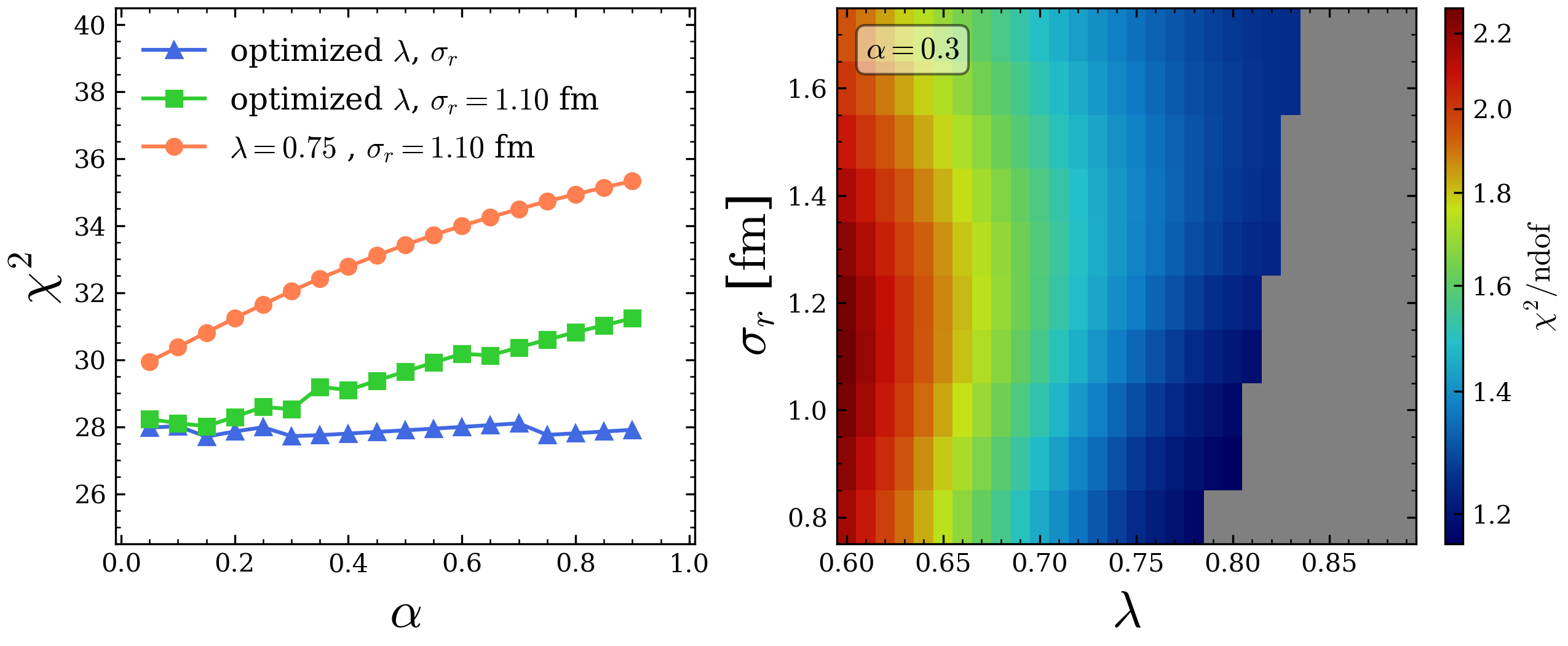}
    \caption{Optimization test on the parameter set ($\alpha, \sigma_r, \lambda$) in MEM-RL algorithm. The left panel shows $\chi^2$ as a function of $\alpha \in (0,1)$ with fixed and optimal choices of $\sigma_r$ and $\lambda$. Right panel shows $\chi^2/\mathrm{dof}$ contour of $\sigma_r, \lambda$ with fixed value of $\alpha=0.3$. Note unphysical results are depicted as gray. The Gaussian source function with $ R=3.5$ fm and $\lambda=0.8$ is used.
    } 
    \label{fig:mem_rl_test}
\end{figure}

\subsection{Test using the MEM-RL algorithm}

\begin{figure*}
    \centering
    \medskip
    \includegraphics[width=\textwidth]{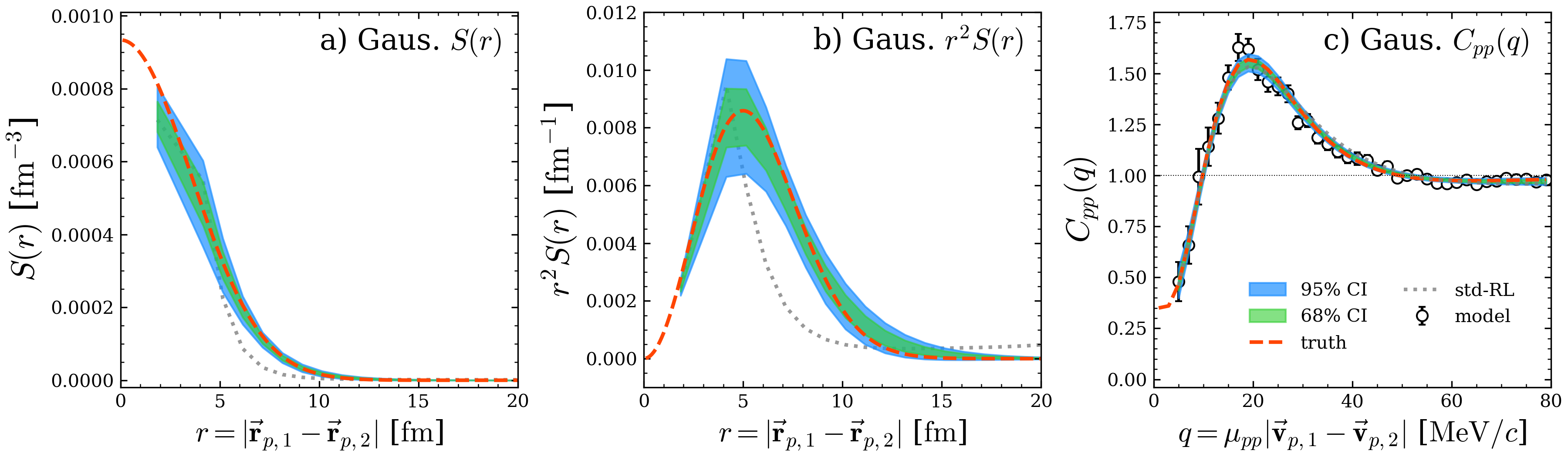}
    \caption{ Results of MEM-RL applied to correlation function calculated from a Gaussian source function. Panel (a) shows the restored source with $1\sigma$ (green) and $2\sigma$ (blue) uncertainty with the truth (red dotted). Panel (b) shows the restored source function scaled by $r^2$ to magnify the comparison at the tail. Panel (c) shows the predicted correlation function in comparison to the generated data. As a reference, the solution from the standard RL algorithm, which converges to an unphysical result, is plotted in a gray dotted line.}
    \label{fig:restored_proton_gaus}
\end{figure*}

First, a few bins of the correlations at small $q < 5~\mathrm{MeV}/c$ are explicitly removed since they are usually absent in real data or have large errors due to acceptance and limited statistics. This blurred $C(q)$, which serves as the input to the MEM-RL algorithm, is shown in panel c of Fig.~\ref{fig:restored_proton_gaus}. 

Our objective is not to recover a source that precisely mirrors the original noise-free correlation but rather to ascertain the probability distribution of the source function that aligns with the blurred data. Employing the MEM regularization necessitates pre-determination of $\alpha$ and $\sigma_r$. The width of the smearing kernel $\sigma_r$ is connected to our prior knowledge of the smoothness of the source function. The regularization strength $\alpha$ acts as a trade-off between the data-driven term and the entropy term in the objective function. In other words, data of better quality (smaller statistical errors) allows adopting smaller values of $\alpha$. 

Along with the purity $\lambda$, these parameters are searched in uniform grids, with the optimal set determined by minimizing the $\chi^2$ of the restored correlation function to the data. However, over-fitting cannot be completely avoided in a certain region of parameter space. For instance, $\lambda$ larger than the true value can only yield a physical source for a certain combination of $\alpha$ and $\sigma_r$.

In practice, a restored source is considered to be unphysical and discarded from the optimization process if it exhibits multiple peaks of $r^2S(r)$ or does not have a vanishing tail at large $r$. The minimum of the parameters for fixed $\alpha=0.3$ is found to lie at the margin at $\lambda=0.8$ and $\sigma_r=0.9$ fm. 

The left panel of Fig.~\ref{fig:mem_rl_test} shows the $\chi^2$ as a function of $\alpha$ with fixed and optimal choices of $\sigma_r$, and $\lambda$. While small $\alpha$ is favored for fixed value of $\lambda$ and $\sigma_r$ (red), it exhibits rough independence on $\alpha$ once $\sigma_r$ and $\lambda$ are both optimized (blue). This suggests that in practice, it suffices to only optimize $\sigma_r$ and $\lambda$ for better efficiency. It is important to note that while an optimal set of hyperparameters exists, there is no definitive metric for determining such a set. 

However, over-fitting is still possible for a certain combination of parameters. For a given $\alpha$, $\lambda$ larger than the true value results in a stronger source at small $r$. To preserve the normalization, the algorithm is forced to introduce unphysical features at larger $r$ unless $\sigma_r$ is increased to suppress the change. Such unphysical sources, as represented by the gray area in the right panel of Fig.~\ref{fig:mem_rl_test}, are explicitly removed from the optimization process.

In all MEM restoration, the convergence criteria are set by considering the mutual cancellation of the data-driven $\Delta^H \psi$ and regularization terms $\Delta^\mathbb{S} \psi$. Specifically, we consider the algorithm to be converged if the relative update drops below a threshold, i.e. $\forall r_j$, 
\begin{gather}
    \label{eq:convergence}
    t_j = \dfrac{|\Delta^H\mathcal{S}_j + \Delta^{\mathbb{S}}\mathcal{S}_j|}{|\Delta^H\mathcal{S}_j|+|\Delta^{\mathbb{S}}\mathcal{S}_j|} < 10^{-3},
\end{gather}
which is usually satisfied in less than 30k iterations. Faster convergence can be understood by realizing that for $\alpha > 0$, the MEM estimates do not have to maximize the likelihood of the data as in the standard RL. 

The uncertainty of the restored source is quantified by the bootstrap method. Given the blurred $C(q)$, we generate an ensemble of data by sampling from a normal distribution with a width determined by the error bars of the correlation, i.e., $\mathcal{N}(C(q), \delta C(q))$. For each sample, the MEM-RL algorithm was applied to restore the source with an optimized parameter set ($\lambda, \alpha, \sigma_r$). The uncertainty of the restored source is then estimated by the standard deviation of the ensemble.

Next, we test the effectiveness of MEM-RL algorithm with the described optimization strategy in recovering source shape and size. We apply the algorithm to a model correlation function associated with a Gaussian source function represented by the open circles in panel~c of Fig.~\ref{fig:restored_proton_gaus}. The parameters of the source function are chosen to be $\lambda_G = 0.65$ and $R_G = 2.5$ fm, respectively. \color{black} Restored sources with $1\sigma$ (green) and $2\sigma$ (blue) uncertainties are shown in panel a and b of Fig.~\ref{fig:restored_proton_gaus}, respectively. While the mean of the source function scaled by $r^2$ is in excellent agreement with the true source function, the restored source deviates slightly from the truth at a small $r$ due to the discretization of kernel and noise in the data. 

One measure of the source size is the FWHM of the source function. For a Gaussian source, the relation between $r_{1/2}$ and $R_G$ is given by $r_{1/2} = 2\sqrt{\ln 2} R_G$. For $R_G = 2.5$ fm, $r_{1/2}^{(\mathrm{true})}=4.16$ fm. The estimated size of the restored source is found to be $r_{1/2}^{(\mathrm{RL})} = 4.25 \pm 0.61$ fm, compared to $r_{1/2}^{(\mathrm{fit})} = 4.20 \pm 0.12$ fm obtained from a Gaussian fit (orange dotted).

In addition to the perturbation introduced to the data, the discrepancy between the true and restored source size can be attributed to the binning. Since the data are binned with the width of $2~\mathrm{MeV}/c$ bin in $q \in (0,80)$ MeV/c, the $r$ dependence of the kernel utilized in the restoration can at most contain the same number of data points, that is, $38$ (minus 2 since we explicitly remove the first 2 bins), to satisfy the requirement of the RL algorithm. The purity parameter, on the other hand, is found to be consistent with the true value $\lambda^{(\mathrm{RL})} = 0.64 \pm 0.04$, which can be observed from the agreement of the predicted correlation function with the data at small $q$ in panel~c of Fig.~\ref{fig:restored_proton_gaus}.

\section{Result and discussion}
\label{sec:result}
\begin{figure*}
    \centering
    \medskip
    \includegraphics[width=\textwidth]{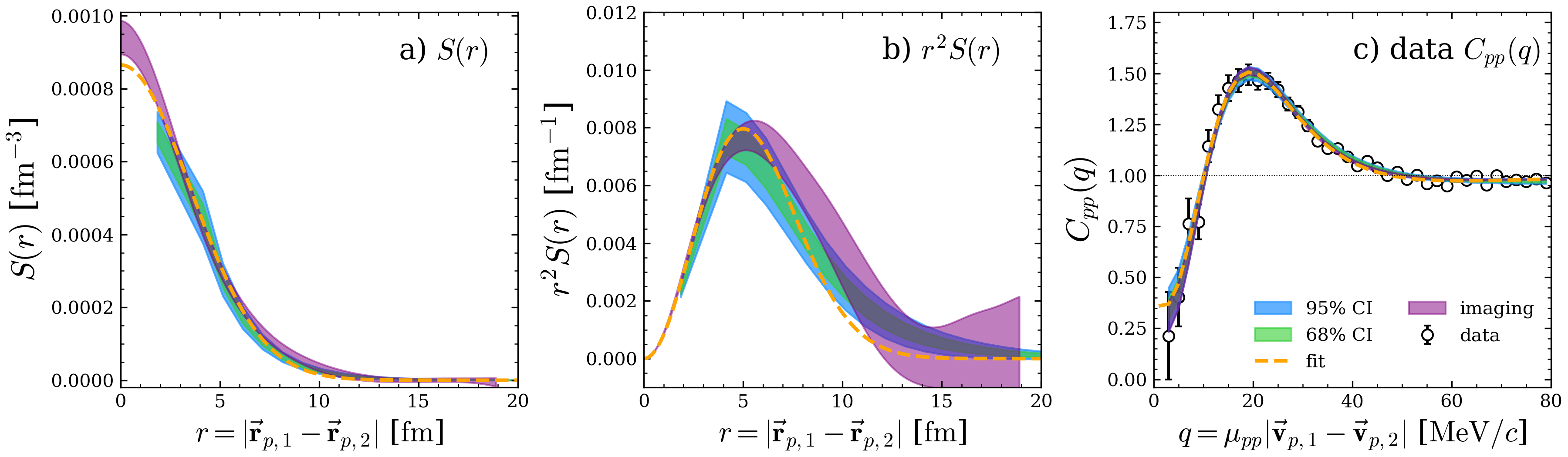}
    \caption{ Results of MEM-RL applied to experimental correlation function~\cite{henzl2011hira}. Panels (a) and (b) show the restored source function and that scaled by $r^2$. Panel (c) shows the predicted correlation function and experimental data. In all panels, the orange dotted line represents the source function by performing $\chi^2$-fit with a Gaussian source function. Green and blue shades represent the result of MEM-RL with $1\sigma$ and $2\sigma$ uncertainties, respectively. As a baseline, results from Bayesian imaging~\cite{henzl2011hira} are plotted in purple shades.}
    \label{fig:restored_proton_hira}
\end{figure*}
In this section, we present the recovered source function from the experimental data. For the two-proton correlations, we consider the experimental data for central, symmetric collisions $^{40}\mathrm{Ca}+^{40}\mathrm{Ca}$ at 80 $\mathrm{MeV}/\mathrm{nucleon}$~\cite{henzl2011hira}. In this study, we chose the correlation function at high total momentum of the pair ($740-900$ MeV/c) and forward angle in the lab frame $\theta_{\mathrm{lab}} = (33^\circ,58^\circ)$. The data is binned with $2\mathrm{MeV}/c$ bin width in the range $q \in (0, 80)\mathrm{MeV}/c$, as depicted in the lower-right panel of Fig.~\ref{fig:restored_proton_hira}. This enables us to leverage a larger number of bins in the position space and consequently facilitates the deblurring of a smoother source function. 

\begin{table}[htbp]
    \centering
    \renewcommand{\arraystretch}{1.4}
    \begin{tabular}{@{}llll@{}}
    \toprule
    {Parameter} & {MEM-RL} & {Gaussian fit} & {Imaging} \\
    \midrule
    $\lambda$          & $0.66^{+0.05}_{-0.05}$ & $0.61^{+0.11}_{-0.08}$ & $0.69^{+0.19}_{-0.12}$ \\
    $r_{1/2}$ [fm]     & $4.17 ^{+0.38}_{-0.38}$  & $4.20^{+0.29}_{-0.21}$ & $4.06^{+0.23}_{-0.40}$ \\
    $\sigma_r$ [fm]    & $1.07 ^{+0.24}_{-0.24}$  & N/A & N/A \\
    \bottomrule
    \end{tabular}
    \caption{Fit parameters for $C_{pp}$ from the experimental data~\cite{henzl2011hira}.}
    \label{tab:experimental_data}
\end{table}

As shown in panels a and b of Fig.~\ref{fig:restored_proton_hira}, the restored source function is consistent with the Gaussian source (orange dotted) obtained from the $\chi^2$ fit~\cite{henzl2011hira}. Additionally, we also compare our results to the Bayesian imaging (purple shades) reported in the same paper. Qualitatively, the peak of $r^2S(r)$ is in good agreement with the Gaussian source, and the tail is more extended, which is consistent with imaging. Quantitatively, the FWHM estimated from MEM-RL is $r_{1/2}^{(RL)}=4.17 \pm 0.38$ fm, as compared to $4.20^{+0.29}_{-0.21}$ fm in the fitted Gaussian source and $4.06^{+0.23}_{-0.40}$ fm obtained from imaging. The extracted purity parameter is also consistent. The results are summarized in the Table~\ref{tab:experimental_data}.

\begin{figure*}
    \centering
    \includegraphics[width=\textwidth]{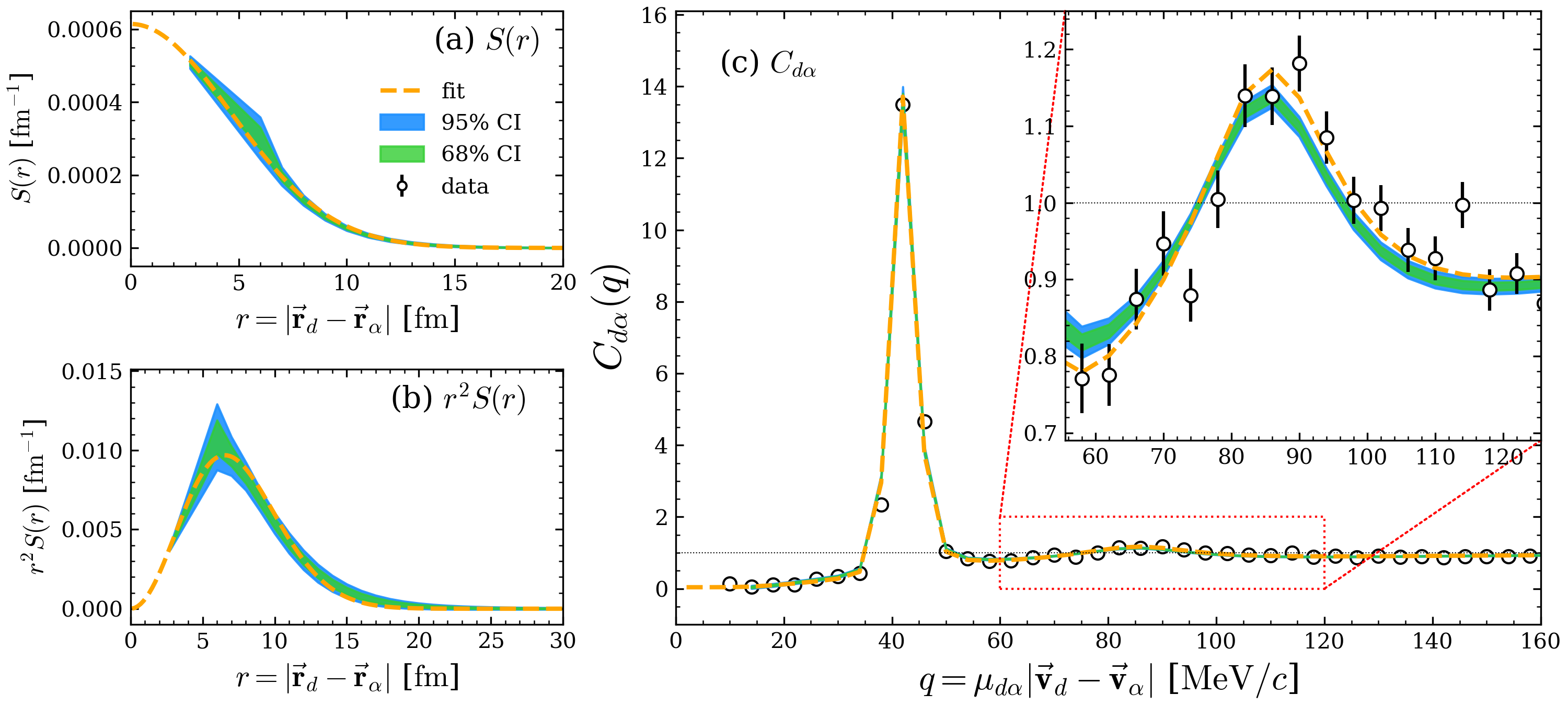}
    
    \caption{Results of the MEM-RL algorithm (blue and green bands) and $\chi^2$-fit (orange dotted) applied to the $C_{d\alpha}$ from HiRA10 data. Panels (a) and (b) show the restored source function and that scaled by $r^2$. The bottom panel (c) shows the predicted correlation functions and data with the second peak zoomed in.}
    \label{fig:restored_dalpha}
\end{figure*}

Next, we extend the algorithm to non-identical pairs. We choose to recover the source function from $d$--$\alpha$ pairs in $^{40}\mathrm{Ca}+^{112}\mathrm{Sn}$ at $140$ $\mathrm{MeV}/\mathrm{nucleon}$~\cite{sweany2021hira}. Unlike the case of $C_{pp}$, $C_{\mathrm{d}\alpha}$ exhibits a sharp peak at $q\approx 42\,\mathrm{MeV}/c$ and a broad peak centered at $q\approx 84\,\mathrm{MeV}/c$, as shown in panel c of Fig.~\ref{fig:restored_dalpha}. The sharp peak is attributed to the decay of the first excited state of $^{6}\mathrm{Li}$ at $\sim 2.186\,\mathrm{MeV}$ ($J^\pi=3^+$) with a width of $\Gamma=0.024\,\mathrm{MeV}$ ($\Gamma_c/\Gamma=1.0$). The broad peak originates from the resonance at $E^*(^{6}\mathrm{Li})\approx4.312\,\mathrm{MeV}$ ($J^\pi=2^+, \Gamma=1.30\,\mathrm{MeV}$) with small contribution from the resonance at $\sim 5.65\,\mathrm{MeV}$ ($J^\pi=1^+, \Gamma=1.9\,\mathrm{MeV}$). At small relative momentum $q$, d$\alpha$ correlation function exhibits large anti-correlation due to the mutual Coulomb interaction. Unlike the tail of pp correlation function, the d$\alpha$ correlation function has not yet reached unity at large $q\gtrsim 250\,\mathrm{MeV}/c$. This is partly due to a stronger Coulomb repulsion and the fact that correlation arising from global conservation laws can not be easily subtracted~\cite{zibi2008conser}. 

As shown in the bottom panel of Fig.~\ref{fig:kernel}, the width of the first resonance peak at $q=42$ MeV/$c$ used when calculating the kernel is very narrow ($<1$MeV/$c$). Thus, the width of that peak in the correlation function is mostly dominated by the resolution of reconstructed charged particles in the experiment - something that the theoretical calculations of the correlation function do not account for. To introduce the detector resolution, the forward model is modified by smearing the kernel with a Gaussian function with width $\sigma_q$. This effect primarily affects the amplitude and width of the first peak and is expected to be moderately independent of the rest of the parameters. The rest of the procedure is identical to that applied to the proton-proton correlation function and described in Section~\ref{sec:formalism}. The employed $d$--$\alpha$ kernel is consistent with that in~\cite{pierre2023deblurring, nzabahimana2025source}. It was calculated using complex Woods-Saxon potentials with an energy-dependent imaginary part \cite{nzabahimana2023particle} fitted to $d$--$\alpha$ phase shift measurements~\cite{boal1990phaseshift}.

We fixed $\alpha=0.3$ and we optimized the rest of the parameters to be $\lambda_{RL} = 0.98 \pm 0.02$, $\sigma_r=1.27\pm 0.13$ $\mathrm{fm}$ and $\sigma_q=2.2$ $\mathrm{MeV}/c$. The estimated source size is $r_{1/2}^{\mathrm{RL}} = 5.39\pm 0.52$ fm compared to $5.44\pm0.06$ fm from Gaussian fit. Fig.~\ref{fig:restored_dalpha} shows that the correlation function is well reproduced by the restored source function. It is worth mentioning that the current framework does not support the consideration of the collective motion where momentum-space correlation in the source function $S\equiv S(q,r)$ significantly affects the shape of the correlation function~\cite{giuseppe2007dalpha}. Consequently, the Koonin Pratt equation~\eqref{eq:KooninPratt} is modified by taking the Hadamard product of the source function and the kernel, meaning the mathematical structure is not applicable to the RL algorithm. 

There are some challenges in applying the method of deblurring to extract the source function from the experimental two-particle correlations. 
Firstly, it is not uncommon to see larger error bars at small or large values of $q$ where the tail of the correlation function is assumed to converge to unity. However, as a result of the low data statistics at those bins, the tail of the correlation function does not converge to unity (e.g., due to the phase-space effect~\cite{zibi2008conser, zibi2009conser}). Since those effects are not accommodated in the kernel, the algorithm would force the source function to attain an unphysical form, such as multiple peaks in $r^2S(r)$, to be able to explain the data even with the strong smoothing regularization. 

Secondly, kernels with interesting structures pose challenges in the binning of the source function. Specifically for $d-\alpha$ correlations (see the bottom of Fig.~\ref{fig:kernel}), that issue introduces uncertainty at either the second peak or the tail. Our solution was to use a customized binning in $r$. An uneven binning was chosen (coarse binning at low $r$ with finer binning at larger values of $r$) such that the calculations using the Koonin-Pratt equation (Eq.~\eqref{eq:KooninPratt}) are approximately the same as those using a continuous source distribution. 

Thirdly, extracting one meaningful number that characterizes the entire source distribution is challenging by itself, independently of the deblurring method. The most common way to do this is to find the value of the FWHM from the source distribution, $r_{1/2}$, described in Sec.~\ref{sec:Test}. However, due to the $r^2$-dependence in the Koonin-Pratt equation, the source function at small $r$ can never be accurately restored because $r^2S(r)\rightarrow 0$ regardless of the value of $S(r=0)$ that is essential to calculate FWHM. In this study, source sizes are estimated by extrapolating to the origin before calculating the FWHM, which contributes to the uncertainty of the extracted parameters of the source.

\section{Summary and outlook}
In this work, we have demonstrated the effectiveness of the MEM-RL algorithm in restoring source functions from the two-particle correlation functions. We have presented the formalism of the algorithm designed specifically for the Koonin-Pratt equation. Unlike TV regularization, we showed that MEM-RL does not suffer from the risk of altering the integral of the source function and is thus more suitable for the application to the correlation function. The method also provides a more flexible way to control the smoothness of the source function in the cost of optimizing an additional parameter in the matrix $\Pi$, compared to TV regularization.  

To facilitate the application in practice, we demonstrated that the value of the $\chi^2$ is moderately independent of the regularization strength $\alpha$ as long as the $\sigma_r$ and $\lambda$ are optimized. As an illustration, we have applied the procedures to both Gaussian $C_{pp}$ and real data. In both cases, essential structures of the source function are restored. In particular, the extracted purity parameters are comparable to the true values and obtained from fits using the Gaussian source and Bayesian imaging. The extracted source sizes from the MEM-RL algorithm are also consistent with the previously published results~\cite{henzl2011hira}. To demonstrate the robustness of our method, we have applied the MEM-RL algorithm to the preliminary $d$--$\alpha$ correlation function from $^{40}\mathrm{Ca}+^{112}\mathrm{Sn}$ at $140$ $\mathrm{MeV}/\mathrm{nucleon}$~\cite{sweany2021hira}. To explain the sharp peak, we have introduced the effect of detector resolution by smearing the kernel with a Gaussian function. The extracted source size, purity, and resolution are consistent with the Gaussian fit. 

We have also discussed the potential challenges and sources of uncertainties in the algorithm's application. Among the issues, the most prominent one is the choice of binning in the position space of the kernel, especially for kernels with sharp features, e.g. for systems like $d$--$\alpha$. 

\section{Acknowledgments}

We would like to acknowledge support from the National Science Foundation (Grant Nos. PHY-2110218 and PHY-1712832). Also, the work was supported by the U.S. Department of Energy through Los Alamos National Laboratory. Los Alamos National Laboratory is operated by Triad National Security, LLC, for the National Nuclear Security Administration of the U.S. Department of Energy~( Contract No. 89233218CNA000001).  Finally, support was provided by the U.S.\ Department of Energy Office of Science under Grant DE-SC0019209.

\bibliography{ref.bib}
\bibliographystyle{apsrev4-2}
\end{document}